\documentclass[aps,pra,reprint,showkeys]{revtex4-2}
\usepackage{graphicx}

\begin{document}

\title{Piezoelectric properties of II--IV/I--V and II--IV/III--III \\
ferroelectric perovskite superlattices}

% repeat the \author .. \affiliation  etc. as needed
% \email, \thanks, \homepage, \altaffiliation all apply to the current
% author. Explanatory text should go in the []'s, actual e-mail
% address or url should go in the {}'s for \email and \homepage.
% Please use the appropriate macro foreach each type of information

% \affiliation command applies to all authors since the last
% \affiliation command. The \affiliation command should follow the
% other information
% \affiliation can be followed by \email, \homepage, \thanks as well.
\author{Alexander I. Lebedev}
\email[]{swan@scon155.phys.msu.ru}
%\homepage[]{Your web page}
%\thanks{}
%\altaffiliation{}
\affiliation{Physics Department, Moscow State University, 119991 Moscow, Russia}

%Collaboration name if desired (requires use of superscriptaddress
%option in \documentclass). \noaffiliation is required (may also be
%used with the \author command).
%\collaboration can be followed by \email, \homepage, \thanks as well.
%\collaboration{}
%\noaffiliation

\date{\today}

\begin{abstract}
The stability of high-symmetry $P4mm$ polar phase in eleven ferroelectric perovskite superlattices
with the polar discontinuity is studied from first principles. In most superlattices, this phase
exhibits either the ferroelectric or the antiferrodistortive instability, or both of them. The
structure of the ground state and, for a number of systems, also of metastable phases in these
superlattices is found. The spontaneous polarization and piezoelectric properties of superlattices
are calculated. The appearance of high piezoelectric coefficients (up to 150--270~pC/N) in some
superlattices is associated with the strain-induced local rearrangement of certain atomic groups
in the primitive cell.

\texttt{Ferroelectrics 567, 89 (2020); DOI: 10.1080/00150193.2020.1791592}

\end{abstract}

% insert suggested keywords - APS authors don't need to do this

%\maketitle must follow title, authors, abstract, and keywords
\maketitle

\section{Introduction}

In recent years, much attention has been paid to studies of low-dimensional structures in which
new physical phenomena that have no analogues in bulk materials have been discovered. Due to
these new functionalities, these materials are considered as very promising for future applications
in electronics. Ferroelectric superlattices (SL)---quasi-two-dimensional structures with artificial
periodicity whose properties can be easily tuned to obtain necessary functionality---belong to this
interesting class of materials.

Most of the previous studies of ferroelectric perovskite superlattices were carried out on
II--IV/II--IV or I--V/I--V type superlattices (here, the numbers indicate the valence of atoms that
enter the $A$ and $B$~sites of the \emph{AB}O$_3$ perovskite structure). In such SLs, there are no
double electric layers at the interface between two dielectric materials, and therefore, they are
macroscopically electrically neutral both in the bulk and at the interface (see
\cite{RevModPhys.77.1083,HandbookChap134,CurrOpinSolidStateMaterSci.12.55,JComputTheorNanosci.5.2071,
PhysSolidState.51.2324,PhysSolidState.52.1448,PhysStatusSolidiB.249.789,PhysSolidState.54.1026,
PhysSolidState.55.1198,PhysSolidState.57.486} and references therein).

Studies of the SrTiO$_3$/LaAlO$_3$ heterostructures have revealed new interesting phenomena that
appear in these structures as a result of the so-called polar discontinuity---of a polarization
jump produced by a violation in the sequence of charged layers at the interface between II--IV and
III--III materials. These effects include the appearance of a conducting layer near the interface
(a two-dimensional electron gas), its magnetism, and even superconductivity~\cite{Nature.427.423,
NatureMater.6.493,Science.317.1196}. These phenomena can be controlled using an external electric
field~\cite{Nature.456.624}. The divergence of the electrostatic potential in such heterostructures,
which results in the appearance of conducting layers at the interface between two dielectrics, was
called a polar catastrophe. The possibility of the appearance of the two-dimensional electron gas
at the interface between a ferroelectric and a nonpolar dielectric in perovskites was systematically
studied in Ref.~\cite{SciRep.6.34667}. Later it was realized that conducting layers can also be
obtained in ferroelectric structures without the polar discontinuity~\cite{PhysRevB.92.115406}.
This made it possible to create new types of electronic devices---ferroelectric structures with
switchable giant tunneling electroresistance~\cite{ApplPhysLett.107.232902,PhysRevB.94.155420},
whose idea is based on earlier works~\cite{Science.313.181,NanoLett.9.427}. We note that the
appearance of similar phenomena can be expected in epitaxial films of II--IV perovskites grown on
DyScO$_3$, GdScO$_3$, NdScO$_3$, and NdGaO$_3$ substrates, in which a polar discontinuity at the
interface is possible.

Until now, theoretical studies of superlattices with the polar discontinuity have focused on the
study of the polarization and electric field distributions in these structures and on the search
for conditions of the appearance of a two-dimensional electron gas at the
interface~\cite{PhysRevB.79.100102,PhysRevLett.103.016804,PhysRevB.80.045425,PhysRevB.80.165130,
PhysRevB.80.241103,PhysRevB.81.235112,PhysRevB.87.085305}. The questions about the stability of
the high-symmetry structure in such SLs, the possibility of phase transitions in them, and the
physical properties of possible low-symmetry phases were not analyzed. At the same time, the
ferroelectric and antiferrodistortive (AFD) instabilities characteristic of many perovskites
can lead to strong distortions of the structure of SLs, and the earlier predictions of physical
properties obtained without taking these distortions into account may be incorrect.

One of the questions that have not been studied in detail earlier is the question about the
piezoelectric properties of superlattices with the polar discontinuity. The calculations of the
properties of such SLs were limited to the calculations of them in the high-symmetry $P4mm$
phase~\cite{JApplPhys.109.066107,ChinPhysB.27.027701}. It is known that the record-high
piezoelectric coefficients in PbTiO$_3$-based solid solutions near the morphotropic
boundary~\cite{JApplPhys.82.1804} are associated with the ease of inclination of the polarization
vector from the [100] direction to the [111] direction under the influence of an electric
field~\cite{PhysRevLett.84.5423}. The polar discontinuity in superlattices enables, under certain
conditions, to create high values of irreversible polarization in their structures. That is why
it seemed interesting to check whether it is possible to obtain high piezoelectric coefficients
in superlattices with the polar discontinuity using the strain-induced inclination of the
polarization vector. In Ref.~\cite{JApplPhys.109.066107}, superlattices with the polar discontinuity
were already considered as a way to obtain stable, weakly temperature-dependent piezoelectric
properties.

\section{Calculation technique}

\begin{figure}
\includegraphics[scale=1.5]{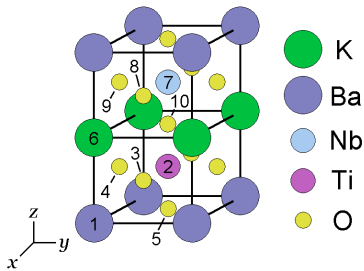}
\caption{\label{fig1}The superlattice geometry.}
\end{figure}

In this work, the properties of eleven free-standing [001]-oriented short-period II--IV/I--V and
II--IV/III--III superlattices with the polar discontinuity and a thickness of individual layers of
one unit cell are studied using first-principles calculations. The geometry of the superlattices
is shown in Fig.~\ref{fig1}. The choice of free-standing superlattices is due to the fact that for
any short-period superlattice grown on a substrate, mismatch dislocations and other defects appear
in the transition layer with a typical thickness of $\sim$100~{\AA} because of a mismatch between
the equilibrium in-plane lattice parameter of the superlattice and that of the substrate. As a result,
the in-plane lattice parameter of the superlattice relaxes to that of the free-standing superlattice.
This is why in most thick experimentally grown SLs, the in-plane lattice parameter is close to that
of a free-standing SL.

The first-principles calculations were performed within the density functional theory using the
\texttt{ABINIT} program and norm-conserving pseudopotentials constructed according to the RRKJ
scheme~\cite{PhysRevB.41.1227} in the local density approximation (LDA), like in
Ref.~\cite{PhysSolidState.51.362}. The cutoff energy was 30~Ha (816~eV) except for Ta-containing
systems, in which it was 40~Ha (1088~eV). Integration over the Brillouin zone was carried out using
a 8$\times$8$\times$4 Monkhorst--Pack mesh for the high-symmetry structure and meshes with
equivalent density of ${\bf k}$-points for the low-symmetry phases. The equilibrium lattice
parameters and atomic positions were calculated by relaxing forces acting on the atoms to values
less than $2 \cdot 10^{-6}$~Ha/Bohr (0.1~meV/{\AA}). The phonon spectra, the tensor of piezoelectric
stress coefficients~$e_{i\mu}$, and the elastic compliance tensor~$S_{\mu\nu}$ were calculated using
the density-functional perturbation theory. The $e_{i\mu}$ values were then converted to piezoelectric
strain coefficients~$d_{i\nu}$ using the formula $d_{i\nu} = e_{i\mu}S_{\mu\nu}$, and for monoclinic
cells the tensor components were transformed to the standard setting of the monoclinic cell, in which
the polarization vector lies in the $xz$~plane.

\section{Results and discussion}

\begin{figure}
\includegraphics{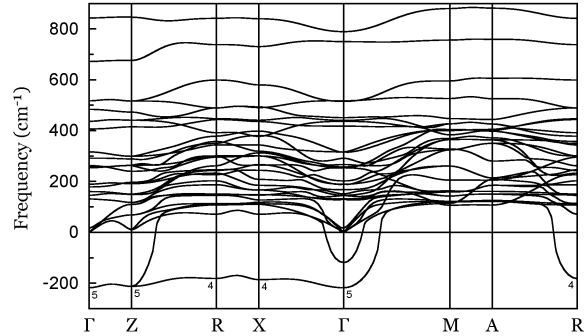}
\caption{\label{fig2}The phonon spectrum of the KNbO$_3$/BaTiO$_3$ superlattice in the $P4mm$~phase.
The numbers near the curves indicate the symmetry of unstable modes.}
\end{figure}

To begin with, it should be noted that the electrical conductivity is usually a factor that prevents
many applications of the ferroelectric properties. Since the experiments on structures with the
polar discontinuity often revealed metallic conductivity at the interface, it was necessary to make
sure that the superlattices under study are insulating. The calculations confirmed that in all
superlattices studied in this work, the conduction band is separated from the valence band by a
sufficiently large energy gap, and so all studied SLs are dielectrics.

Since the layer sequence in considered superlattices does not admit the reversal of $ z \to -z$
(Fig.~\ref{fig1}), the superlattices are always polar and their high-symmetry phase has the $P4mm$
symmetry. However, this structure can exhibit various instabilities characteristic of crystals with
the perovskite structure: either the ferroelectric instability, or the antiferrodistortive one, or
both of them simultaneously. The ground state of superlattices was searched in the traditional
way~\cite{PhysSolidState.51.2324,PhysSolidState.52.1448}: first, the structures resulting from the
condensation of all unstable phonons found in the phonon spectrum of the $P4mm$ phase were
calculated, taking into account their possible degeneracy. Then, by calculating the phonon spectra
at all high-symmetry points of the Brillouin zone and the elastic tensor of these structures, the
stability of the obtained solutions was checked. In the case when an instability in any of these
structures was found, the search for the ground state was continued until a structure whose phonon
spectrum has no unstable modes and the positive definite matrix composed of the elastic tensor
components in the Voigt notation is found. In this case, the conclusion that a stable phase which
may be a ground state can be made. The problem here is in that in chains of phases generated by
different octahedra rotations, several stable states can be found, as was shown recently for
SrTiO$_3$~\cite{PhysSolidState.58.300}. In this case, the ground state is the stable phase with
the lowest total energy, and other stable phases should be considered as metastable. If the energy
of such metastable solutions differs little from the energy of the ground state, they should be
considered as solutions that can be observed in experiment, and for them, as well as for the ground
state, an analysis of their physical properties should also be carried out.

Calculations of the phonon spectra showed that in most studied SLs (except for BaTiO$_3$/LaAlO$_3$,
SrTiO$_3$/LaAlO$_3$, PbTiO$_3$/KTaO$_3$, and PbTiO$_3$/LaAlO$_3$), the high-symmetry $P4mm$ phase
exhibits the ferroelectric instability with respect to the in-plane distortion of the structure
or, in other words, with respect to the inclination of the polarization vector. The phonon spectrum
of a such superlattice, KNbO$_3$/BaTiO$_3$, is shown in Fig.~\ref{fig2}. It can be seen that in
addition to the instability at the $\Gamma$ point, the instabilities also appear at the $X$, $R$,
and $Z$ points of the Brillouin zone. We have already encountered a similar situation in
KNbO$_3$/KTaO$_3$~\cite{PhysStatusSolidiB.249.789} and BaTiO$_3$/BaZrO$_3$~\cite{PhysSolidState.55.1198}
SLs.

An instability zone, which is observed as a band of imaginary phonon frequencies on the
$\Gamma$--$Z$--$R$--$X$-$\Gamma$ line (imaginary frequencies are represented in the figure by negative
numbers), is a consequence of the ferroelectric instability in ...--Ti--O--... chains propagating in
the plane of SL--the so-called chain instability~\cite{PhysRevLett.74.4067}. Indeed, an analysis of
phonon eigenvectors of these modes shows that at all above-mentioned points of the Brillouin zone,
the out-of-phase, transverse, $xy$-polarized displacements of Ti and O atoms in chains propagating
in the [100] and [010] directions dominate in the vibrations. In superlattices based on BaTiO$_3$ and
SrTiO$_3$, the displacements of the Nb(Ta) atoms were small, whereas in superlattices based on
SrZrO$_3$ and BaZrO$_3$, as well as in the PbTiO$_3$/KNbO$_3$ SL, the chain instability appears
mainly in ...--Nb--O--... chains. At the center of the Brillouin zone, the described displacements
pattern corresponds to a doubly degenerate ferroelectric $E$~mode ($\Gamma_5$).%
\footnote{The numbers of irreducible representations used in this work follow their classification
adopted at Bilbao Crystallographic Server~\cite{Bilbao}.}

\begin{table*}
\caption{\label{table1}The energies (in meV per 10-atom supercell) of different low-symmetry phases resulting
from condensation of unstable phonons at different points of the Brillouin zone for KNbO$_3$/BaTiO$_3$
and KNbO$_3$/SrZrO$_3$ short-period superlattices.}
\begin{ruledtabular}
\begin{tabular}{cccccc}
Phase & Unstable phonon & Energy & Phase & Unstable phonon & Energy \\
\hline
\multicolumn{6}{c}{KNbO$_3$/BaTiO$_3$ superlattice} \\
$P4mm$ & ---   & 0       & $Pm$     & $\Gamma_5 (\eta,0)$    & $-$27.8 \\
$Abm2$ & $R_4$ & $-$13.8 & $Cmc2_1$ & $Z_5 (\eta,\eta)$      & $-$31.0 \\
$Pma2$ & $X_4$ & $-$14.8 & $Cm$     & $\Gamma_5 (\eta,\eta)$ & $-$38.6 \\
$Pmc2_1$ & $Z_5 (\eta,0)$ & $-$21.3 \\
\multicolumn{6}{c}{KNbO$_3$/SrZrO$_3$ superlattice} \\
$P4mm$ & ---   & 0       & $Pc$   & $Z_5 (\eta,0) + \Gamma_5 (0,\xi)$ & $-$19.2 \\
$I4cm$ & $A_4$ & $-$1.3  & $Pma2$ & $M_5 (\eta,0)$                    & $-$19.9 \\
$P4bm$ & $M_4$ & $-$2.1  & $Cm$   & $\Gamma_5 (\eta,\eta)$            & $-$20.0 \\
$Abm2$ & $R_4$ & $-$4.7  & $P2$   & $X_3 + X_4$                       & $-$20.7 \\
$Pma2$ & $X_4$ & $-$6.7  & $Cc$   & $A_4 + \Gamma_5 (\eta,\eta)$      & $-$24.1 \\
$Pmm2$ & $X_3$ & $-$12.4 & $Pm$   & $X_3 + \Gamma_5 (0,\eta)$         & $-$24.9 \\
$Pmc2_1$ & $Z_5 (\eta,0)$          & $-$12.9 & $Cmm2$ & $M_5 (\eta,\eta)$                   & $-$26.9 \\
$Pm$   & $\Gamma_5 (\eta,0)$       & $-$13.7 & $Pm$   & $X_3 + \Gamma_5 (\eta,0)$           & $-$27.6 \\
$Pc$   & $R_4 + \Gamma_5 (\eta,0)$ & $-$15.4 & $Pm$   & $M_5 (\eta,0) + \Gamma_5 (\xi,\xi)$ & $-$36.9 \\
$Pc$   & $X_4 + \Gamma_5 (\eta,0)$ & $-$17.0 & $Cm$   & $M_4 + \Gamma_5 (\eta,0)$           & $-$39.4 \\
$Cmc2_1$ & $Z_5 (\eta,\eta)$       & $-$18.9 & $Pc$   & $M_4 + \Gamma_5 (\eta,\eta)$        & $-$51.9 \\
\end{tabular}
\end{ruledtabular}
\end{table*}

Of two possible polar phases resulting from the condensation of the unstable $E$ mode, the $Cm$
phase with atomic displacements along the [110] direction was the lowest-energy phase for all
superlattices. The displacements in the [100] direction were energetically less favorable, which
is apparently due to a tendency of bulk BaTiO$_3$ and KNbO$_3$ to polarize along the [111] direction
in the ground state. As an example, the energies of different low-symmetry phases for the
KNbO$_3$/BaTiO$_3$ SL are given in Table~\ref{table1}. The structures obtained from the condensation
of $Z_5$, $R_4$, and $X_4$ phonons had a higher energy as compared to that of the $Cm$ phase.

The absence of the ferroelectric instability in the $P4mm$ phase of PbTiO$_3$/KTaO$_3$ and
PbTiO$_3$/LaAlO$_3$ superlattices can be explained by a tendency of bulk PbTiO$_3$ to polarize
along the [001] axis. This instability was also absent in the SrTiO$_3$/LaAlO$_3$ SL, in which
both constituents are nonpolar. In addition, the ferroelectric instability did not appear in the
BaTiO$_3$/LaAlO$_3$ superlattice, in which its absence is a consequence of the strong (by 2.3\%)
in-plane compression of BaTiO$_3$ layers. Our calculations of the effect of strain on the
ground-state structure of BaTiO$_3$ showed that the biaxial compression of 1\% is sufficient to
change the most stable polar phase in it to the $P4mm$ phase, in agreement
with~\cite{PhysRevLett.80.1988,PhysRevB.69.212101}.

\begin{figure}
\includegraphics{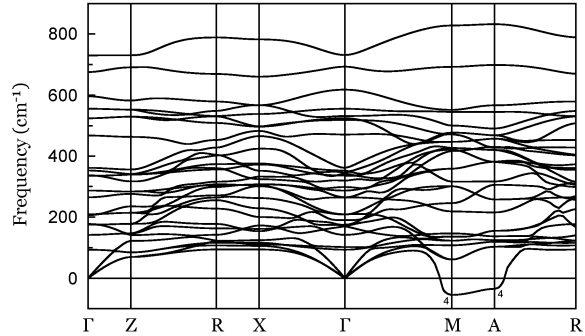}
\caption{\label{fig3}The phonon spectrum of the BaTiO$_3$/LaAlO$_3$ superlattice in the
$P4mm$ phase. The numbers near the curves indicate the symmetry of unstable modes.}
\end{figure}

Along with the ferroelectric instability, in the high-symmetry $P4mm$ phase of a number of
superlattices (KNbO$_3$/SrTiO$_3$, KNbO$_3$/SrZrO$_3$, BaTiO$_3$/LaAlO$_3$, SrTiO$_3$/KTaO$_3$,
SrTiO$_3$/LaAlO$_3$, and PbTiO$_3$/LaAlO$_3$) an AFD instability with the octahedra rotations
around the $z$~axis is observed. It can be clearly seen by the appearance of unstable phonons on
the $M$--$A$ line (Fig.~\ref{fig3}). The occurrence of the AFD instability in SLs clearly
correlates with the existence of this instability in one or both of the constituent materials
(SrTiO$_3$, SrZrO$_3$, LaAlO$_3$). A comparison of the energies of phases resulting from the
condensation of phonons at the $M$ and $A$~points of the Brillouin zone shows that of two phases
resulting from the $M_4$ phonon condensation (space group $P4bm$) and from the $A_4$ phonon
condensation (space group $I4cm$), the $P4bm$ phase was always more energetically favorable.
However, since the energy difference between these phases is small, and each of them can exhibit
the ferroelectric instability, it was necessary to consider all polar subgroups of these phases
when searching for the ground state.

The calculations showed that the ferroelectric instability of the $P4bm$ and $I4cm$ phases
is characteristic of KNbO$_3$/SrTiO$_3$, SrTiO$_3$/KTaO$_3$, and KNbO$_3$/SrZrO$_3$ SLs. Of two
structures, $Cm$ and $Pc$, into which the $P4bm$ structure can transform upon polar distortion,
the $Pc$ phase with the polarization along the [110] direction of the pseudocubic cell always had
a lower energy. For a random starting polar displacement from the $I4cm$ phase, the structure
always relaxed to the $Cc$ phase, in which the polarization is also directed along the [110]
direction of the pseudocubic cell. For first two superlattices, the energy difference between the
$Cc$ and $Pc$ phases was only 0.5--0.7~meV. Taking into account that the analysis proved the
stability of both phases, the $Cc$ phase should be considered as metastable. As the energy
difference between the two phases is very small, there is a real possibility that both structures
can occur in the experiment simultaneously. That is why the properties of these superlattices were
calculated below for both the $Pc$ and $Cc$ phases.

\begin{table*}
\caption{\label{table2}Calculated polarization and the energy gain upon the ferroelectric
distortion for all studied short-period superlattices with the polar discontinuity. The
polarization values are in C/m$^2$, the energy values are in meV.}
\begin{ruledtabular}
\begin{tabular}{cccccc}
Superlattice & Space group & $P_x$ & $P_y$ & $P_z$ & $\Delta E$ \\
\hline
KNbO$_3$/PbTiO$_3$  & $Cm$ & 0.181 & 0 & 0.567 & 3.34 \\
KNbO$_3$/BaTiO$_3$  & $Cm$ & 0.338 & 0 & $-$0.033 & 38.61 \\
KNbO$_3$/BaZrO$_3$  & $Cm$ & 0.229 & 0 & 0.181 & 33.82 \\
KNbO$_3$/SrTiO$_3$  & $Pc$ & 0.281 & 0 & 0.066 & 53.31; 8.96\footnotemark[1] \\
KNbO$_3$/SrZrO$_3$  & $Pc$ & 0.268 & 0 & 0.187 & 51.86; 49.80\footnotemark[1] \\
BaTiO$_3$/KTaO$_3$  & $Cm$ & 0.180 & 0 & 0.130 & 9.73 \\
BaTiO$_3$/LaAlO$_3$ & $P4bm$ & 0   & 0 & 0.057 & --- \\
SrTiO$_3$/KTaO$_3$  & $Pc$ & 0.105 & 0 & 0.141 & 11.86; 0.82\footnotemark[1] \\
SrTiO$_3$/LaAlO$_3$ & $P4bm$ & 0   & 0 & 0.012 & --- \\
PbTiO$_3$/KTaO$_3$  & $P4mm$ & 0   & 0 & 0.454 & --- \\
PbTiO$_3$/LaAlO$_3$ & $P4bm$ & 0   & 0 & 0.119 & --- \\
\end{tabular}
\footnotetext[1]{The energies relative to that of the ``nonpolar'' $P4bm$~phase.}
\end{ruledtabular}
\end{table*}

The most complex picture was observed in superlattices, in which both ferroelectric and AFD
instabilities were simultaneously present. An example of such a system is the KNbO$_3$/SrZrO$_3$
SL. The energies of different phases for this superlattice are given in Table~\ref{table1}.

In addition to the above-discussed unstable $M_4$ mode, one more doubly degenerate unstable
$M_5$~mode was observed in the phonon spectra of the $P4mm$ phase of four SLs (KNbO$_3$/SrTiO$_3$,
KNbO$_3$/SrZrO$_3$, SrTiO$_3$/LaAlO$_3$, and PbTiO$_3$/LaAlO$_3$). This mode describes the rotations
of the octahedra around one or both of the $x$ and $y$~axes. The distortions described by this mode
with the ($\eta$,0) and ($\eta$,$\eta$) order parameters resulted in the $Pma2$ and $Cmm2$ phases,
of which the $Cmm2$~phase always had a lower energy. Both these phases are characterized by the
ferroelectric instability. An analysis of the polar subgroups of the $Pma2$ and $Cmm2$ phases in
KNbO$_3$/SrTiO$_3$ and KNbO$_3$/SrZrO$_3$ SLs showed that among them the $Pc$ phase, which is
already familiar to us, has the lowest energy. It may seem strange that the same $Pc$ phase appears
as a result of condensation of phonons described by two different irreducible representations
($M_4$ and $M_5$). However, it should be taken into account that the vertical axis of the octahedra
in the $Pc$~phase is slightly inclined, that is, in reality this phase is described by two nonzero
rotations around the coordinate axes. This explains why two structures, in which the distortions are
described by different irreducible representations, relax to the same $Pc$~phase when the polar
displacements are switched on. In SrTiO$_3$/LaAlO$_3$ and PbTiO$_3$/LaAlO$_3$ SLs, both $Pma2$ and
$Cmm2$ structures relax to the ``nonpolar'' $P4bm$ phase when the polar displacements are switched on.

The space groups of the energetically most favorable phases obtained for all studied SLs are given
in Table~\ref{table2}. It is seen that in superlattices exhibiting only the ferroelectric instability,
the $Cm$ phase is the ground state. In SLs exhibiting only the AFD instability, the $P4bm$ phase is
the ground state. And, finally, in superlattices in which both instabilities are present in the $P4mm$
phase, the $Pc$ phase is the ground state. The difference between our results and the results of
earlier calculations for the KNbO$_3$/BaTiO$_3$ SL~\cite{PhysRevB.87.085305} is due to the fact that
the calculations~\cite{PhysRevB.87.085305} were performed for the superlattice clamped on the
SrTiO$_3$ substrate: in such SL, the $P4mm$ phase is indeed the ground state.

The energy gain $\Delta E$ per 10-atom formula unit, which results from the ferroelectric distortion,
is also given in Table~\ref{table2}. The obtained values show that at room temperature, the
predicted ground-state structures are likely to be observed for KNbO$_3$/BaTiO$_3$, KNbO$_3$/BaZrO$_3$,
and KNbO$_3$/SrZrO$_3$ SLs. For the ground-state structures of all superlattices, the spontaneous
polarization and the piezoelectric tensor will be calculated below.

Classical electrostatics of an electrically neutral interface between two dielectrics requires that
the components of the electric displacement field normal to this interface are equal in two materials.
If the materials have different spontaneous polarizations, then a bound electric charge appears at
the interface, and the electric field generated by it makes the electric displacement fields equal
in two materials.

In superlattices with the polar discontinuity, the violation of the order of charged $A$O and $B$O$_2$
planes generates an additional electrostatic perturbation at the interfaces and creates a polarization
jump of $\Delta P = e/2a^2$ at every interface in the superlattices (here $a$ is the in-plane lattice
parameter of the SL)~\cite{PhysRevB.79.100102}. An elegant solution to this problem, which can be used
in the general case, was proposed in~\cite{PhysRevB.80.241103} within the framework of the modern
theory of polarization (the Berry phase formalism). An application of this approach to our superlattices
enabled us to calculate the electric displacement field in them and to use it to determine an average
polarization of SLs. In this work, we are interested precisely in this quantity. The polarization values
in individual layers can be calculated by correcting the obtained average polarization taking into
account the jump in the ionic contribution to the Berry phase at the interface and dielectric constants
of individual constituents.%
\footnote{The periodicity of a superlattice assumes that the electric field strengths $E_1$
and $E_2$ in its layers satisfy the condition $E_1x_1 + E_2x_2 = 0$, where $x_1$ and $x_2$
are the thicknesses of individual layers. This condition combined with the equation
$(P_1 + \epsilon_1 E_1) - (P_2 + \epsilon_2 E_2) = \Delta P$, in which $P_1$, $P_2$,
$\epsilon_1$, and $\epsilon_2$ are the spontaneous polarizations and dielectric constants in
two layers, gives a solution to this problem in the linear approximation.}

When calculating the polarization by the Berry phase method, it should be borne in mind that the ionic
contributions to the Berry phase are different in nonpolar $Pm{\bar 3}m$ phases of II--IV and I--V
perovskites. This is why for each SL with the polar discontinuity it is necessary first to find the
Berry phase of a nonpolar structure before calculating the polarization. Unfortunately, in our case
the determination of the $z$~component of the Berry phase is a problem because in the high-symmetry
$P4mm$ phase, which does not have a mirror plane $\sigma_z$, it is impossible to reverse the
polarization or to construct a non-polar structure. To \emph{estimate} $P_z$, we considered unrelaxed
structures with ideal atomic positions corresponding to the cubic perovskite structure in both layers.
In this structure, the electron contribution to the Berry phase is nonzero because of the
redistribution of the electron density between the layers, and the ionic contribution reflects the
difference in the Berry phases of individual perovskites. The change of the Berry phase upon the
transition from the described unrelaxed structure to the ground-state structure was used to calculate
the average polarization using the standard formula. To correctly determine the polarization from the
change of the Berry phase, which is determined with an accuracy of $2\pi m$, for all SLs the
calculations were also performed for one intermediate point at which the atoms are located halfway
between the unrelaxed structure and the ground state.

\begin{table*}
\caption{\label{table3}Nonzero components of the piezoelectric tensor $d_{i\nu}$
(in pC/N) in the ground state of all studied short-period superlattices with the
polar discontinuity.}
\begin{ruledtabular}
\begin{tabular}{ccccccccccc}
Superlattice & $d_{11}$ & $d_{12}$ & $d_{13}$ & $d_{15}$ & $d_{24}$ & $d_{26}$ & $d_{31}$ & $d_{32}$ & $d_{33}$ & $d_{35}$ \\
\hline
KNbO$_3$/PbTiO$_3$  & 13.4 & 3.7  & $-$13.9 & 158.3 & 273.3  & 137.3 & $-$3.0  & $-$2.9  & $-$10.9 & $-$0.8 \\
KNbO$_3$/BaTiO$_3$  & 23.5 & 9.6  & $-$12.7 & 3.8   & $-$2.8 & 106.4 & 10.7    & 7.9     & $-$16.3 & 17.4 \\
KNbO$_3$/BaZrO$_3$  & 16.3 & 4.7  & $-$11.7 & 5.6   & 7.9    & 20.8  & $-$8.9  & $-$7.3  & 16.7    & 1.8 \\
KNbO$_3$/SrTiO$_3$  & 35.9 & 15.3 & $-$19.1 & 8.3   & 41.7   & 151.0 & $-$22.0 & $-$12.1 & 16.1    & 9.4 \\
KNbO$_3$/SrZrO$_3$  & 30.8 & 10.1 & $-$18.8 & 9.1   & 11.0   & 85.5  & $-$18.1 & $-$9.8  & 17.7    & 12.9 \\
BaTiO$_3$/KTaO$_3$  & 19.2 & 9.5  & $-$18.5 & 2.5   & 7.9    & 25.3  & $-$18.1 & $-$11.6 & 43.4    & 10.1 \\
BaTiO$_3$/LaAlO$_3$ & ---  & ---  & ---     & 1.4   & 1.4    & ---   & 3.0     & 3.0     & $-$4.6  & --- \\
SrTiO$_3$/KTaO$_3$  & 39.7 & 22.8 & $-$27.6 & 28.2  & 46.9   & 41.2  & $-$18.4 & $-$13.0 & 26.1    & $-$4.2 \\
SrTiO$_3$/LaAlO$_3$ & ---  & ---  & ---     & 2.0   & 2.0    & ---   & 1.0     & 1.0     & $-$1.6  & --- \\
PbTiO$_3$/KTaO$_3$  & ---  & ---  & ---     & 91.4  & 91.4   & ---   & $-$8.8  & $-$8.8  & 41.2    & --- \\
PbTiO$_3$/LaAlO$_3$ & ---  & ---  & ---     & 158.6 & 158.6  & ---   & $-$4.3  & $-$4.3  & $-$2.8  & --- \\
\end{tabular}
\end{ruledtabular}
\end{table*}

The calculated polarizations are given in Table~\ref{table2}. The components of the polarization
vector are given relative to the axes of standard crystallographic settings for tetragonal and
monoclinic cells (their axes are rotated in the $xy$ plane by 45$^\circ$ relative to each other).
The comparison of polarizations calculated in structures with and without octahedral rotations shows
that in structures with the $Pc$ space group, the neglect of the AFD rotations can lead to an error
in determining $P_z$ up to 30\% and, in some cases, even to an error in the sign of this quantity.
For the $P4mm$ phases, the obtained values agree well with the published values of
$P_z = 0.532$~C/m$^2$ for the PbTiO$_3$/KNbO$_3$ SL and $P_z = 0.202$~C/m$^2$ for the
PbTiO$_3$/LaAlO$_3$ SL~\cite{ChinPhysB.27.027701}, and with the $P_z = 0.38$~C/m$^2$ value for the
PbTiO$_3$/KTaO$_3$ SL~\cite{ChinPhysLett.33.026302}.

The reason for our interest to the ferroelectric instability in superlattices with the polar
discontinuity is that in such SLs it is possible to obtain sufficiently high piezoelectric coefficients
associated with the ferroelectric phase transitions occurring in them. The literature data on the
piezoelectric properties of such superlattices are limited to calculations for the $P4mm$ phases
of PbTiO$_3$/LaAlO$_3$ and KNbO$_3$/PbTiO$_3$ SLs~\cite{JApplPhys.109.066107,ChinPhysB.27.027701}
and of the PbTiO$_3$/KTaO$_3$ one~\cite{ChinPhysLett.33.026302}. As the ground-state structure in
the first two superlattices differs from $P4mm$, there is a need for more correct calculations for
these superlattices. For other SLs considered in this work, no data on their piezoelectric properties
exist.

In SLs with a tetragonal ground-state structure, in which the polarization is directed along the
$z$~axis, five components of the piezoelectric tensor are nonzero. Among them, the highest values of
$d_{i\nu}$ were obtained for PbTiO$_3$/KTaO$_3$ and PbTiO$_3$/LaAlO$_3$ SLs (Table~\ref{table3}).
Interestingly, among these coefficients, the $d_{15}$ values turned out to be the largest. This
coefficient characterizes the polarization $P_x$ that appears as a result of the $xz$~shear strain
of the unit cell, that is, as a result of inclination of the polarization vector. However, no
clear correlation between $d_{15}$ and $P_z$ values was observed. Moreover, in the related system,
BaTiO$_3$/LaAlO$_3$, the piezoelectric coefficients were very small (Table~\ref{table3}). This means
that the inclination of the polarization vector as a way of obtaining high values of piezoelectric
coefficients is not effective.

In SLs with a monoclinic ground-state structure, in which the polarization vector lies in the
$xz$~plane, the piezoelectric tensor is characterized by ten nonzero components. In these structures,
the highest values of $d_{i\nu}$ were the $d_{24}$ and $d_{26}$ coefficients, which describe the
appearance of polarization in the $y$ direction normal to the $xz$ plane under the $yz$ and $xy$
shear strain. An analysis of the obtained data also does not find a clear correlation between the
piezoelectric coefficients and the average polarization in these structures. The stretching of the
unit cell in the $xz$ plane does result in a change in polarization, but the corresponding
piezoelectric coefficients ($d_{11}$ and $d_{33}$, see Table~\ref{table3}) are not the largest.

\begin{table}
\caption{\label{table4}The values of $\partial u_1^i/\partial \sigma_{15}$ (in {\AA}) of the
$\partial u_{\alpha}^i/\partial \sigma_{\mu\nu}$ tensors for all atoms in the ground-state
$P4bm$ structure of PbTiO$_3$/LaAlO$_3$ and BaTiO$_3$/LaAlO$_3$ superlattices ($A ={}$ Pb and
Ba, respectively). Atoms 11--20 are located in the adjacent cell of the doubled unit cell of
the high-temperature phase.}
\begin{ruledtabular}
\begin{tabular}{ccc}
Atom $i$ & PbTiO$_3$/LaAlO$_3$ & BaTiO$_3$/LaAlO$_3$ \\
\hline
$A$(1) & +5.18    & +0.086 \\
Ti(2)  & +2.37    & $-$0.083 \\
O(3)   & $-$1.19  & +0.239 \\
O(4)   & $-$5.61  & +0.099 \\
O(5)   & $-$1.02  & $-$0.109 \\
La(6)  & +4.56    & +0.731 \\
Al(7)  & +0.76    & +0.016 \\
O(8)   & $-$0.21  & $-$0.314 \\
O(9)   & $-$0.22  & $-$0.154 \\
O(10)  & $-$0.84  & $-$0.097 \\
$A$(11) & +1.71   & $-$0.383 \\
Ti(12)  & +2.37   & $-$0.083 \\
O(13)   & $-$1.19 & +0.239 \\
O(14)   & $-$5.61 & +0.099 \\
O(15)   & $-$1.02 & $-$0.109 \\
La(16)  & +0.48   & +0.370 \\
Al(17)  & +0.76   & +0.016 \\
O(18)   & $-$0.21 & $-$0.314 \\
O(19)   & $-$0.22 & $-$0.154 \\
O(20)   & $-$0.84 & $-$0.097 \\
\end{tabular}
\end{ruledtabular}
\end{table}

To understand the mechanism of the appearance of high piezoelectric coefficients in some SLs with
the polar discontinuity, we analyzed the third-rank tensors $\partial u_{\alpha}^i/\partial \sigma_{\mu\nu}$.
This tensor characterizes the displacement of the $i$th atom in the unit cell in the $\alpha$ direction
produced by the strain described by the $\sigma_{\mu\nu}$ tensor. It turned out that in superlattices
exhibiting the strong piezoelectric response, the values of some components of these tensors for some
atoms reach 10--15~{\AA} (that is, a deformation of the unit cell by 1\% causes the atomic displacements
that exceed 0.1~{\AA}). For example, in the PbTiO$_3$/LaAlO$_3$ SL, such atoms are Pb(1), La(6), and
two oxygen O(4) and O(14) atoms located in the TiO$_2$ layer (Table~\ref{table4}); the contributions
of the O(8) and O(9) oxygen atoms located in the AlO$_2$ layer are 25~times smaller. Isolation of
such atomic groups and analysis of their local structure may be a way to better understand the
microscopic mechanism of the appearance of the strong piezoelectricity and to use this information
for intentional modification of materials in order to obtain high piezoelectric properties.

A comparison of the piezoelectric properties of the metastable $Cc$~phase and the ground-state $Pc$
phase for KNbO$_3$/SrTiO$_3$ and SrTiO$_3$/KTaO$_3$ SLs showed that their piezoelectric tensors are
fairly close to each other, with a typical deviation of the piezoelectric coefficients of 4--6\%.

According to our calculations, the piezoelectric coefficients in studied SLs can reach 150--270~pC/N.
A comparison of the obtained results with the published data finds their reasonable agreement. For
the $P4mm$ phase of the PbTiO$_3$/LaAlO$_3$ SL, our result $e_{33} = -3.52$~C/m$^2$ is close to the
$e_{33} = -2.85$~C/m$^2$ value calculated in~\cite{JApplPhys.109.066107}. However, our
$d_{33} = -18.9$~pC/N value calculated for the same phase of this SL disagrees with the
$d_{33} = +13.9$~pC/N value obtained in~\cite{ChinPhysB.27.027701}. These values are close in magnitude
but different in sign. A possible reason for a stronger discrepancy here may be the neglect
in~\cite{ChinPhysB.27.027701} of the difference in displacements and effective charges of the oxygen
atoms (their effective charge varies from $-$2.07 to $-$5.84). As for the PbTiO$_3$/KTaO$_3$
superlattice studied in~\cite{ChinPhysLett.33.026302}, the values of the $e_{33}$ and $e_{15}$
coefficients obtained there are two orders of magnitude lower than our data, and the $e_{31}$ coefficient
is close to our result in magnitude, but differs in sign.

\section{Conclusions}

Using first-principles calculations, the stability of high-symmetry $P4mm$ polar phase in eleven
ferroelectric perovskite superlattices with the polar discontinuity have been studied. It was shown
that in most superlattices, this phase exhibits either the ferroelectric or the antiferrodistortive
(AFD), or both of these instabilities simultaneously. In superlattices exhibiting only the ferroelectric
instability, the $Cm$ phase is the ground state. In superlattices exhibiting only the AFD instability,
the $P4bm$ phase is the ground state. And finally, in superlattices in which the $P4mm$ phase exhibits
both the ferroelectric and AFD instabilities, the $Pc$ phase is the ground state. In superlattices
whose structure exhibits the AFD instability, the structure of metastable phases was also calculated.
The average spontaneous polarization and piezoelectric properties for the ground-state structures of
all superlattices were calculated. It was shown that the appearance of high piezoelectric coefficients
is due to the strain-induced local rearrangement of certain atomic groups inside the primitive cell.

\mbox{}

\begin{acknowledgments}
This work was supported by the Russian Foundation for Basic Research (RFBR) under Grant 17-02-01068.
\end{acknowledgments}

% Create the reference section using BibTeX:
%\bibliography{all}

\begin{thebibliography}{41}%
\makeatletter
\providecommand \@ifxundefined [1]{%
 \@ifx{#1\undefined}
}%
\providecommand \@ifnum [1]{%
 \ifnum #1\expandafter \@firstoftwo
 \else \expandafter \@secondoftwo
 \fi
}%
\providecommand \@ifx [1]{%
 \ifx #1\expandafter \@firstoftwo
 \else \expandafter \@secondoftwo
 \fi
}%
\providecommand \natexlab [1]{#1}%
\providecommand \enquote  [1]{``#1''}%
\providecommand \bibnamefont  [1]{#1}%
\providecommand \bibfnamefont [1]{#1}%
\providecommand \citenamefont [1]{#1}%
\providecommand \href@noop [0]{\@secondoftwo}%
\providecommand \href [0]{\begingroup \@sanitize@url \@href}%
\providecommand \@href[1]{\@@startlink{#1}\@@href}%
\providecommand \@@href[1]{\endgroup#1\@@endlink}%
\providecommand \@sanitize@url [0]{\catcode `\\12\catcode `\$12\catcode
  `\&12\catcode `\#12\catcode `\^12\catcode `\_12\catcode `\%12\relax}%
\providecommand \@@startlink[1]{}%
\providecommand \@@endlink[0]{}%
\providecommand \url  [0]{\begingroup\@sanitize@url \@url }%
\providecommand \@url [1]{\endgroup\@href {#1}{\urlprefix }}%
\providecommand \urlprefix  [0]{URL }%
\providecommand \Eprint [0]{\href }%
\providecommand \doibase [0]{https://doi.org/}%
\providecommand \selectlanguage [0]{\@gobble}%
\providecommand \bibinfo  [0]{\@secondoftwo}%
\providecommand \bibfield  [0]{\@secondoftwo}%
\providecommand \translation [1]{[#1]}%
\providecommand \BibitemOpen [0]{}%
\providecommand \bibitemStop [0]{}%
\providecommand \bibitemNoStop [0]{.\EOS\space}%
\providecommand \EOS [0]{\spacefactor3000\relax}%
\providecommand \BibitemShut  [1]{\csname bibitem#1\endcsname}%
\let\auto@bib@innerbib\@empty
%</preamble>
\bibitem [{\citenamefont {Dawber}\ \emph {et~al.}(2005)\citenamefont {Dawber},
  \citenamefont {Rabe},\ and\ \citenamefont {Scott}}]{RevModPhys.77.1083}%
  \BibitemOpen
  \bibfield  {author} {\bibinfo {author} {\bibfnamefont {M.}~\bibnamefont
  {Dawber}}, \bibinfo {author} {\bibfnamefont {K.~M.}\ \bibnamefont {Rabe}},\
  and\ \bibinfo {author} {\bibfnamefont {J.~F.}\ \bibnamefont {Scott}},\
  }\bibfield  {title} {\bibinfo {title} {Physics of thin-film ferroelectric
  oxides},\ }\href {https://doi.org/10.1103/RevModPhys.77.1083} {\bibfield
  {journal} {\bibinfo  {journal} {Rev. Mod. Phys.}\ }\textbf {\bibinfo {volume}
  {77}},\ \bibinfo {pages} {1083} (\bibinfo {year} {2005})}\BibitemShut
  {NoStop}%
\bibitem [{\citenamefont {Ghosez}\ and\ \citenamefont
  {Junquera}(2006)}]{HandbookChap134}%
  \BibitemOpen
  \bibfield  {author} {\bibinfo {author} {\bibfnamefont {P.}~\bibnamefont
  {Ghosez}}\ and\ \bibinfo {author} {\bibfnamefont {J.}~\bibnamefont
  {Junquera}},\ }\bibfield  {title} {\bibinfo {title} {First-principle modeling
  of ferroelectric oxide nanostructures},\ }in\ \href@noop {} {\emph {\bibinfo
  {booktitle} {Handbook of Theoretical and Computational Nanotechnology}}},\
  Vol.~\bibinfo {volume} {9},\ \bibinfo {editor} {edited by\ \bibinfo {editor}
  {\bibfnamefont {M.}~\bibnamefont {Rieth}}\ and\ \bibinfo {editor}
  {\bibfnamefont {W.}~\bibnamefont {Schommers}}}\ (\bibinfo  {publisher}
  {American Scientific Publishers},\ \bibinfo {year} {2006})\ pp.\ \bibinfo
  {pages} {623--728}\BibitemShut {NoStop}%
\bibitem [{\citenamefont {Bao}(2008)}]{CurrOpinSolidStateMaterSci.12.55}%
  \BibitemOpen
  \bibfield  {author} {\bibinfo {author} {\bibfnamefont {D.}~\bibnamefont
  {Bao}},\ }\bibfield  {title} {\bibinfo {title} {Multilayered
  dielectric/ferroelectric thin films and superlattices},\ }\href
  {https://doi.org/10.1016/j.cossms.2009.01.006} {\bibfield  {journal}
  {\bibinfo  {journal} {Curr. Opin. Solid State Mater. Sci.}\ }\textbf
  {\bibinfo {volume} {12}},\ \bibinfo {pages} {55} (\bibinfo {year}
  {2008})}\BibitemShut {NoStop}%
\bibitem [{\citenamefont {Junquera}\ and\ \citenamefont
  {Ghosez}(2008)}]{JComputTheorNanosci.5.2071}%
  \BibitemOpen
  \bibfield  {author} {\bibinfo {author} {\bibfnamefont {J.}~\bibnamefont
  {Junquera}}\ and\ \bibinfo {author} {\bibfnamefont {P.}~\bibnamefont
  {Ghosez}},\ }\bibfield  {title} {\bibinfo {title} {First-principles study of
  ferroelectric oxide epitaxial thin films and superlattices: Role of the
  mechanical and electrical boundary conditions},\ }\href
  {https://doi.org/10.1166/jctn.2008.1101} {\bibfield  {journal} {\bibinfo
  {journal} {J. Comput. Theor. Nanosci.}\ }\textbf {\bibinfo {volume} {5}},\
  \bibinfo {pages} {2071} (\bibinfo {year} {2008})}\BibitemShut {NoStop}%
\bibitem [{\citenamefont
  {Lebedev}(2009{\natexlab{a}})}]{PhysSolidState.51.2324}%
  \BibitemOpen
  \bibfield  {author} {\bibinfo {author} {\bibfnamefont {A.~I.}\ \bibnamefont
  {Lebedev}},\ }\bibfield  {title} {\bibinfo {title} {Ab initio studies of
  dielectric, piezoelectric, and elastic properties of BaTiO$_3$/SrTiO$_3$
  ferroelectric superlattices},\ }\href
  {https://doi.org/10.1134/S1063783409110225} {\bibfield  {journal} {\bibinfo
  {journal} {Phys. Solid State}\ }\textbf {\bibinfo {volume} {51}},\ \bibinfo
  {pages} {2324} (\bibinfo {year} {2009}{\natexlab{a}})}\BibitemShut {NoStop}%
\bibitem [{\citenamefont {Lebedev}(2010)}]{PhysSolidState.52.1448}%
  \BibitemOpen
  \bibfield  {author} {\bibinfo {author} {\bibfnamefont {A.~I.}\ \bibnamefont
  {Lebedev}},\ }\bibfield  {title} {\bibinfo {title} {Ground state and
  properties of ferroelectric superlattices based on crystals of the perovskite
  family},\ }\href {https://doi.org/10.1134/S1063783410070218} {\bibfield
  {journal} {\bibinfo  {journal} {Phys. Solid State}\ }\textbf {\bibinfo
  {volume} {52}},\ \bibinfo {pages} {1448} (\bibinfo {year}
  {2010})}\BibitemShut {NoStop}%
\bibitem [{\citenamefont {Lebedev}(2012)}]{PhysStatusSolidiB.249.789}%
  \BibitemOpen
  \bibfield  {author} {\bibinfo {author} {\bibfnamefont {A.~I.}\ \bibnamefont
  {Lebedev}},\ }\bibfield  {title} {\bibinfo {title} {Ground-state structure of
  KNbO$_3$/KTaO$_3$ superlattices: Array of nearly independent
  ferroelectrically ordered planes},\ }\href
  {https://doi.org/10.1002/pssb.201147350} {\bibfield  {journal} {\bibinfo
  {journal} {Phys. Status Solidi B}\ }\textbf {\bibinfo {volume} {249}},\
  \bibinfo {pages} {789} (\bibinfo {year} {2012})}\BibitemShut {NoStop}%
\bibitem [{\citenamefont {Yuzyuk}(2012)}]{PhysSolidState.54.1026}%
  \BibitemOpen
  \bibfield  {author} {\bibinfo {author} {\bibfnamefont {Y.~I.}\ \bibnamefont
  {Yuzyuk}},\ }\bibfield  {title} {\bibinfo {title} {Raman scattering spectra
  of ceramics, films, and superlattices of ferroelectric perovskites: A
  review},\ }\href {https://doi.org/10.1134/S1063783412050502} {\bibfield
  {journal} {\bibinfo  {journal} {Phys. Solid State}\ }\textbf {\bibinfo
  {volume} {54}},\ \bibinfo {pages} {1026} (\bibinfo {year}
  {2012})}\BibitemShut {NoStop}%
\bibitem [{\citenamefont {Lebedev}(2013)}]{PhysSolidState.55.1198}%
  \BibitemOpen
  \bibfield  {author} {\bibinfo {author} {\bibfnamefont {A.~I.}\ \bibnamefont
  {Lebedev}},\ }\bibfield  {title} {\bibinfo {title} {Properties of
  BaTiO$_3$/BaZrO$_3$ ferroelectric superlattices with competing
  instabilities},\ }\href {https://doi.org/10.1134/S1063783413060218}
  {\bibfield  {journal} {\bibinfo  {journal} {Phys. Solid State}\ }\textbf
  {\bibinfo {volume} {55}},\ \bibinfo {pages} {1198} (\bibinfo {year}
  {2013})}\BibitemShut {NoStop}%
\bibitem [{\citenamefont {Tikhonov}\ \emph {et~al.}(2015)\citenamefont
  {Tikhonov}, \citenamefont {Razumnaya}, \citenamefont {Maslova}, \citenamefont
  {Zakharchenko}, \citenamefont {Yuzyuk}, \citenamefont {Ortega}, \citenamefont
  {Kumar},\ and\ \citenamefont {Katiyar}}]{PhysSolidState.57.486}%
  \BibitemOpen
  \bibfield  {author} {\bibinfo {author} {\bibfnamefont {Y.~A.}\ \bibnamefont
  {Tikhonov}}, \bibinfo {author} {\bibfnamefont {A.~G.}\ \bibnamefont
  {Razumnaya}}, \bibinfo {author} {\bibfnamefont {O.~A.}\ \bibnamefont
  {Maslova}}, \bibinfo {author} {\bibfnamefont {I.~N.}\ \bibnamefont
  {Zakharchenko}}, \bibinfo {author} {\bibfnamefont {Y.~I.}\ \bibnamefont
  {Yuzyuk}}, \bibinfo {author} {\bibfnamefont {N.}~\bibnamefont {Ortega}},
  \bibinfo {author} {\bibfnamefont {A.}~\bibnamefont {Kumar}},\ and\ \bibinfo
  {author} {\bibfnamefont {R.~S.}\ \bibnamefont {Katiyar}},\ }\bibfield
  {title} {\bibinfo {title} {Phase transitions in two- and three-component
  perovskite superlattices},\ }\href
  {https://doi.org/10.1134/S1063783415030336} {\bibfield  {journal} {\bibinfo
  {journal} {Phys. Solid State}\ }\textbf {\bibinfo {volume} {57}},\ \bibinfo
  {pages} {486} (\bibinfo {year} {2015})}\BibitemShut {NoStop}%
\bibitem [{\citenamefont {Ohtomo}\ and\ \citenamefont
  {Hwang}(2004)}]{Nature.427.423}%
  \BibitemOpen
  \bibfield  {author} {\bibinfo {author} {\bibfnamefont {A.}~\bibnamefont
  {Ohtomo}}\ and\ \bibinfo {author} {\bibfnamefont {H.~Y.}\ \bibnamefont
  {Hwang}},\ }\bibfield  {title} {\bibinfo {title} {A high-mobility electron
  gas at the LaAlO$_3$/SrTiO$_3$ heterointerface},\ }\href
  {https://doi.org/10.1038/nature02308} {\bibfield  {journal} {\bibinfo
  {journal} {Nature}\ }\textbf {\bibinfo {volume} {427}},\ \bibinfo {pages}
  {423} (\bibinfo {year} {2004})}\BibitemShut {NoStop}%
\bibitem [{\citenamefont {Brinkman}\ \emph {et~al.}(2007)\citenamefont
  {Brinkman}, \citenamefont {Huijben}, \citenamefont {van Zalk}, \citenamefont
  {Huijben}, \citenamefont {Zeitler}, \citenamefont {Maan}, \citenamefont
  {van~der Wiel}, \citenamefont {Rijnders}, \citenamefont {Blank},\ and\
  \citenamefont {Hilgenkamp}}]{NatureMater.6.493}%
  \BibitemOpen
  \bibfield  {author} {\bibinfo {author} {\bibfnamefont {A.}~\bibnamefont
  {Brinkman}}, \bibinfo {author} {\bibfnamefont {M.}~\bibnamefont {Huijben}},
  \bibinfo {author} {\bibfnamefont {M.}~\bibnamefont {van Zalk}}, \bibinfo
  {author} {\bibfnamefont {J.}~\bibnamefont {Huijben}}, \bibinfo {author}
  {\bibfnamefont {U.}~\bibnamefont {Zeitler}}, \bibinfo {author} {\bibfnamefont
  {J.~C.}\ \bibnamefont {Maan}}, \bibinfo {author} {\bibfnamefont {W.~G.}\
  \bibnamefont {van~der Wiel}}, \bibinfo {author} {\bibfnamefont
  {G.}~\bibnamefont {Rijnders}}, \bibinfo {author} {\bibfnamefont {D.~H.~A.}\
  \bibnamefont {Blank}},\ and\ \bibinfo {author} {\bibfnamefont
  {H.}~\bibnamefont {Hilgenkamp}},\ }\bibfield  {title} {\bibinfo {title}
  {Magnetic effects at the interface between non-magnetic oxides},\ }\href
  {https://doi.org/10.1038/nmat1931} {\bibfield  {journal} {\bibinfo  {journal}
  {Nature Mater.}\ }\textbf {\bibinfo {volume} {6}},\ \bibinfo {pages} {493}
  (\bibinfo {year} {2007})}\BibitemShut {NoStop}%
\bibitem [{\citenamefont {Reyren}\ \emph {et~al.}(2007)\citenamefont {Reyren},
  \citenamefont {Thiel}, \citenamefont {Caviglia}, \citenamefont {Kourkoutis},
  \citenamefont {Hammerl}, \citenamefont {Richter}, \citenamefont {Schneider},
  \citenamefont {Kopp}, \citenamefont {R\"uetschi}, \citenamefont {Jaccard},
  \citenamefont {Gabay}, \citenamefont {Muller}, \citenamefont {Triscone},\
  and\ \citenamefont {Mannhart}}]{Science.317.1196}%
  \BibitemOpen
  \bibfield  {author} {\bibinfo {author} {\bibfnamefont {N.}~\bibnamefont
  {Reyren}}, \bibinfo {author} {\bibfnamefont {S.}~\bibnamefont {Thiel}},
  \bibinfo {author} {\bibfnamefont {A.~D.}\ \bibnamefont {Caviglia}}, \bibinfo
  {author} {\bibfnamefont {L.~F.}\ \bibnamefont {Kourkoutis}}, \bibinfo
  {author} {\bibfnamefont {G.}~\bibnamefont {Hammerl}}, \bibinfo {author}
  {\bibfnamefont {C.}~\bibnamefont {Richter}}, \bibinfo {author} {\bibfnamefont
  {C.~W.}\ \bibnamefont {Schneider}}, \bibinfo {author} {\bibfnamefont
  {T.}~\bibnamefont {Kopp}}, \bibinfo {author} {\bibfnamefont {A.-S.}\
  \bibnamefont {R\"uetschi}}, \bibinfo {author} {\bibfnamefont
  {D.}~\bibnamefont {Jaccard}}, \bibinfo {author} {\bibfnamefont
  {M.}~\bibnamefont {Gabay}}, \bibinfo {author} {\bibfnamefont {D.~A.}\
  \bibnamefont {Muller}}, \bibinfo {author} {\bibfnamefont {J.-M.}\
  \bibnamefont {Triscone}},\ and\ \bibinfo {author} {\bibfnamefont
  {J.}~\bibnamefont {Mannhart}},\ }\bibfield  {title} {\bibinfo {title}
  {Superconducting interfaces between insulating oxides},\ }\href
  {https://doi.org/10.1126/science.1146006} {\bibfield  {journal} {\bibinfo
  {journal} {Science}\ }\textbf {\bibinfo {volume} {317}},\ \bibinfo {pages}
  {1196} (\bibinfo {year} {2007})}\BibitemShut {NoStop}%
\bibitem [{\citenamefont {Caviglia}\ \emph {et~al.}(2008)\citenamefont
  {Caviglia}, \citenamefont {Gariglio}, \citenamefont {Reyren}, \citenamefont
  {Jaccard}, \citenamefont {Schneider}, \citenamefont {Gabay}, \citenamefont
  {Thiel}, \citenamefont {Hammerl}, \citenamefont {Mannhart},\ and\
  \citenamefont {Triscone}}]{Nature.456.624}%
  \BibitemOpen
  \bibfield  {author} {\bibinfo {author} {\bibfnamefont {A.~D.}\ \bibnamefont
  {Caviglia}}, \bibinfo {author} {\bibfnamefont {S.}~\bibnamefont {Gariglio}},
  \bibinfo {author} {\bibfnamefont {N.}~\bibnamefont {Reyren}}, \bibinfo
  {author} {\bibfnamefont {D.}~\bibnamefont {Jaccard}}, \bibinfo {author}
  {\bibfnamefont {T.}~\bibnamefont {Schneider}}, \bibinfo {author}
  {\bibfnamefont {M.}~\bibnamefont {Gabay}}, \bibinfo {author} {\bibfnamefont
  {S.}~\bibnamefont {Thiel}}, \bibinfo {author} {\bibfnamefont
  {G.}~\bibnamefont {Hammerl}}, \bibinfo {author} {\bibfnamefont
  {J.}~\bibnamefont {Mannhart}},\ and\ \bibinfo {author} {\bibfnamefont
  {J.-M.}\ \bibnamefont {Triscone}},\ }\bibfield  {title} {\bibinfo {title}
  {Electric field control of the LaAlO$_3$/SrTiO$_3$ interface ground state},\
  }\href {https://doi.org/10.1038/nature07576} {\bibfield  {journal} {\bibinfo
  {journal} {Nature}\ }\textbf {\bibinfo {volume} {456}},\ \bibinfo {pages}
  {624} (\bibinfo {year} {2008})}\BibitemShut {NoStop}%
\bibitem [{\citenamefont {Yang}\ \emph {et~al.}(2016)\citenamefont {Yang},
  \citenamefont {Nazir}, \citenamefont {Behtash},\ and\ \citenamefont
  {Cheng}}]{SciRep.6.34667}%
  \BibitemOpen
  \bibfield  {author} {\bibinfo {author} {\bibfnamefont {K.}~\bibnamefont
  {Yang}}, \bibinfo {author} {\bibfnamefont {S.}~\bibnamefont {Nazir}},
  \bibinfo {author} {\bibfnamefont {M.}~\bibnamefont {Behtash}},\ and\ \bibinfo
  {author} {\bibfnamefont {J.}~\bibnamefont {Cheng}},\ }\bibfield  {title}
  {\bibinfo {title} {High-throughput design of two-dimensional electron gas
  systems based on polar/nonpolar perovskite oxide heterostructures},\ }\href
  {https://doi.org/10.1038/srep34667} {\bibfield  {journal} {\bibinfo
  {journal} {Sci. Rep.}\ }\textbf {\bibinfo {volume} {6}},\ \bibinfo {pages}
  {34667} (\bibinfo {year} {2016})}\BibitemShut {NoStop}%
\bibitem [{\citenamefont {Yin}\ \emph {et~al.}(2015)\citenamefont {Yin},
  \citenamefont {Aguado-Puente}, \citenamefont {Qu},\ and\ \citenamefont
  {Artacho}}]{PhysRevB.92.115406}%
  \BibitemOpen
  \bibfield  {author} {\bibinfo {author} {\bibfnamefont {B.}~\bibnamefont
  {Yin}}, \bibinfo {author} {\bibfnamefont {P.}~\bibnamefont {Aguado-Puente}},
  \bibinfo {author} {\bibfnamefont {S.}~\bibnamefont {Qu}},\ and\ \bibinfo
  {author} {\bibfnamefont {E.}~\bibnamefont {Artacho}},\ }\bibfield  {title}
  {\bibinfo {title} {Two-dimensional electron gas at the PbTiO$_3$/SrTiO$_3$
  interface: An ab initio study},\ }\href
  {https://doi.org/10.1103/PhysRevB.92.115406} {\bibfield  {journal} {\bibinfo
  {journal} {Phys. Rev. B}\ }\textbf {\bibinfo {volume} {92}},\ \bibinfo
  {pages} {115406} (\bibinfo {year} {2015})}\BibitemShut {NoStop}%
\bibitem [{\citenamefont {Ruan}\ \emph {et~al.}(2015)\citenamefont {Ruan},
  \citenamefont {Qiu}, \citenamefont {Yuan}, \citenamefont {Ji}, \citenamefont
  {Wang}, \citenamefont {Li},\ and\ \citenamefont
  {Wu}}]{ApplPhysLett.107.232902}%
  \BibitemOpen
  \bibfield  {author} {\bibinfo {author} {\bibfnamefont {J.}~\bibnamefont
  {Ruan}}, \bibinfo {author} {\bibfnamefont {X.}~\bibnamefont {Qiu}}, \bibinfo
  {author} {\bibfnamefont {Z.}~\bibnamefont {Yuan}}, \bibinfo {author}
  {\bibfnamefont {D.}~\bibnamefont {Ji}}, \bibinfo {author} {\bibfnamefont
  {P.}~\bibnamefont {Wang}}, \bibinfo {author} {\bibfnamefont {A.}~\bibnamefont
  {Li}},\ and\ \bibinfo {author} {\bibfnamefont {D.}~\bibnamefont {Wu}},\
  }\bibfield  {title} {\bibinfo {title} {Improved memory functions in
  multiferroic tunnel junctions with a dielectric/ferroelectric composite
  barrier},\ }\href {https://doi.org/10.1063/1.4937390} {\bibfield  {journal}
  {\bibinfo  {journal} {Appl. Phys. Lett.}\ }\textbf {\bibinfo {volume}
  {107}},\ \bibinfo {pages} {232902} (\bibinfo {year} {2015})}\BibitemShut
  {NoStop}%
\bibitem [{\citenamefont {Wu}\ \emph {et~al.}(2016)\citenamefont {Wu},
  \citenamefont {Shen}, \citenamefont {Yang}, \citenamefont {Zhou},
  \citenamefont {Chen},\ and\ \citenamefont {Feng}}]{PhysRevB.94.155420}%
  \BibitemOpen
  \bibfield  {author} {\bibinfo {author} {\bibfnamefont {Q.}~\bibnamefont
  {Wu}}, \bibinfo {author} {\bibfnamefont {L.}~\bibnamefont {Shen}}, \bibinfo
  {author} {\bibfnamefont {M.}~\bibnamefont {Yang}}, \bibinfo {author}
  {\bibfnamefont {J.}~\bibnamefont {Zhou}}, \bibinfo {author} {\bibfnamefont
  {J.}~\bibnamefont {Chen}},\ and\ \bibinfo {author} {\bibfnamefont {Y.~P.}\
  \bibnamefont {Feng}},\ }\bibfield  {title} {\bibinfo {title} {Giant tunneling
  electroresistance induced by ferroelectrically switchable two-dimensional
  electron gas at nonpolar BaTiO$_3$/SrTiO$_3$ interface},\ }\href
  {https://doi.org/10.1103/PhysRevB.94.155420} {\bibfield  {journal} {\bibinfo
  {journal} {Phys. Rev. B}\ }\textbf {\bibinfo {volume} {94}},\ \bibinfo
  {pages} {155420} (\bibinfo {year} {2016})}\BibitemShut {NoStop}%
\bibitem [{\citenamefont {Tsymbal}\ and\ \citenamefont
  {Kohlstedt}(2006)}]{Science.313.181}%
  \BibitemOpen
  \bibfield  {author} {\bibinfo {author} {\bibfnamefont {E.~Y.}\ \bibnamefont
  {Tsymbal}}\ and\ \bibinfo {author} {\bibfnamefont {H.}~\bibnamefont
  {Kohlstedt}},\ }\bibfield  {title} {\bibinfo {title} {Tunneling across a
  ferroelectric},\ }\href {https://doi.org/10.1126/science.1126230} {\bibfield
  {journal} {\bibinfo  {journal} {Science}\ }\textbf {\bibinfo {volume}
  {313}},\ \bibinfo {pages} {181} (\bibinfo {year} {2006})}\BibitemShut
  {NoStop}%
\bibitem [{\citenamefont {Velev}\ \emph {et~al.}(2009)\citenamefont {Velev},
  \citenamefont {Duan}, \citenamefont {Burton}, \citenamefont {Smogunov},
  \citenamefont {Niranjan}, \citenamefont {Tosatti}, \citenamefont {Jaswal},\
  and\ \citenamefont {Tsymbal}}]{NanoLett.9.427}%
  \BibitemOpen
  \bibfield  {author} {\bibinfo {author} {\bibfnamefont {J.~P.}\ \bibnamefont
  {Velev}}, \bibinfo {author} {\bibfnamefont {C.-G.}\ \bibnamefont {Duan}},
  \bibinfo {author} {\bibfnamefont {J.~D.}\ \bibnamefont {Burton}}, \bibinfo
  {author} {\bibfnamefont {A.}~\bibnamefont {Smogunov}}, \bibinfo {author}
  {\bibfnamefont {M.~K.}\ \bibnamefont {Niranjan}}, \bibinfo {author}
  {\bibfnamefont {E.}~\bibnamefont {Tosatti}}, \bibinfo {author} {\bibfnamefont
  {S.~S.}\ \bibnamefont {Jaswal}},\ and\ \bibinfo {author} {\bibfnamefont
  {E.~Y.}\ \bibnamefont {Tsymbal}},\ }\bibfield  {title} {\bibinfo {title}
  {Magnetic tunnel junctions with ferroelectric barriers: Prediction of four
  resistance states from first principles},\ }\href
  {https://doi.org/10.1021/nl803318d} {\bibfield  {journal} {\bibinfo
  {journal} {Nano Lett.}\ }\textbf {\bibinfo {volume} {9}},\ \bibinfo {pages}
  {427} (\bibinfo {year} {2009})}\BibitemShut {NoStop}%
\bibitem [{\citenamefont {Murray}\ and\ \citenamefont
  {Vanderbilt}(2009)}]{PhysRevB.79.100102}%
  \BibitemOpen
  \bibfield  {author} {\bibinfo {author} {\bibfnamefont {{\'E}.~D.}\
  \bibnamefont {Murray}}\ and\ \bibinfo {author} {\bibfnamefont
  {D.}~\bibnamefont {Vanderbilt}},\ }\bibfield  {title} {\bibinfo {title}
  {Theoretical investigation of polarization-compensated II-IV/I-V perovskite
  superlattices},\ }\href {https://doi.org/10.1103/PhysRevB.79.100102}
  {\bibfield  {journal} {\bibinfo  {journal} {Phys. Rev. B}\ }\textbf {\bibinfo
  {volume} {79}},\ \bibinfo {pages} {100102} (\bibinfo {year}
  {2009})}\BibitemShut {NoStop}%
\bibitem [{\citenamefont {Niranjan}\ \emph {et~al.}(2009)\citenamefont
  {Niranjan}, \citenamefont {Wang}, \citenamefont {Jaswal},\ and\ \citenamefont
  {Tsymbal}}]{PhysRevLett.103.016804}%
  \BibitemOpen
  \bibfield  {author} {\bibinfo {author} {\bibfnamefont {M.~K.}\ \bibnamefont
  {Niranjan}}, \bibinfo {author} {\bibfnamefont {Y.}~\bibnamefont {Wang}},
  \bibinfo {author} {\bibfnamefont {S.~S.}\ \bibnamefont {Jaswal}},\ and\
  \bibinfo {author} {\bibfnamefont {E.~Y.}\ \bibnamefont {Tsymbal}},\
  }\bibfield  {title} {\bibinfo {title} {Prediction of a switchable
  two-dimensional electron gas at ferroelectric oxide interfaces},\ }\href
  {https://doi.org/10.1103/PhysRevLett.103.016804} {\bibfield  {journal}
  {\bibinfo  {journal} {Phys. Rev. Lett.}\ }\textbf {\bibinfo {volume} {103}},\
  \bibinfo {pages} {016804} (\bibinfo {year} {2009})}\BibitemShut {NoStop}%
\bibitem [{\citenamefont {Bristowe}\ \emph {et~al.}(2009)\citenamefont
  {Bristowe}, \citenamefont {Artacho},\ and\ \citenamefont
  {Littlewood}}]{PhysRevB.80.045425}%
  \BibitemOpen
  \bibfield  {author} {\bibinfo {author} {\bibfnamefont {N.~C.}\ \bibnamefont
  {Bristowe}}, \bibinfo {author} {\bibfnamefont {E.}~\bibnamefont {Artacho}},\
  and\ \bibinfo {author} {\bibfnamefont {P.~B.}\ \bibnamefont {Littlewood}},\
  }\bibfield  {title} {\bibinfo {title} {Oxide superlattices with alternating
  $p$ and $n$ interfaces},\ }\href {https://doi.org/10.1103/PhysRevB.80.045425}
  {\bibfield  {journal} {\bibinfo  {journal} {Phys. Rev. B}\ }\textbf {\bibinfo
  {volume} {80}},\ \bibinfo {pages} {045425} (\bibinfo {year}
  {2009})}\BibitemShut {NoStop}%
\bibitem [{\citenamefont {Wang}\ \emph {et~al.}(2009)\citenamefont {Wang},
  \citenamefont {Niranjan}, \citenamefont {Jaswal},\ and\ \citenamefont
  {Tsymbal}}]{PhysRevB.80.165130}%
  \BibitemOpen
  \bibfield  {author} {\bibinfo {author} {\bibfnamefont {Y.}~\bibnamefont
  {Wang}}, \bibinfo {author} {\bibfnamefont {M.~K.}\ \bibnamefont {Niranjan}},
  \bibinfo {author} {\bibfnamefont {S.~S.}\ \bibnamefont {Jaswal}},\ and\
  \bibinfo {author} {\bibfnamefont {E.~Y.}\ \bibnamefont {Tsymbal}},\
  }\bibfield  {title} {\bibinfo {title} {First-principles studies of a
  two-dimensional electron gas at the interface in ferroelectric oxide
  heterostructures},\ }\href {https://doi.org/10.1103/PhysRevB.80.165130}
  {\bibfield  {journal} {\bibinfo  {journal} {Phys. Rev. B}\ }\textbf {\bibinfo
  {volume} {80}},\ \bibinfo {pages} {165130} (\bibinfo {year}
  {2009})}\BibitemShut {NoStop}%
\bibitem [{\citenamefont {Stengel}\ and\ \citenamefont
  {Vanderbilt}(2009)}]{PhysRevB.80.241103}%
  \BibitemOpen
  \bibfield  {author} {\bibinfo {author} {\bibfnamefont {M.}~\bibnamefont
  {Stengel}}\ and\ \bibinfo {author} {\bibfnamefont {D.}~\bibnamefont
  {Vanderbilt}},\ }\bibfield  {title} {\bibinfo {title} {Berry-phase theory of
  polar discontinuities at oxide-oxide interfaces},\ }\href
  {https://doi.org/10.1103/PhysRevB.80.241103} {\bibfield  {journal} {\bibinfo
  {journal} {Phys. Rev. B}\ }\textbf {\bibinfo {volume} {80}},\ \bibinfo
  {pages} {241103} (\bibinfo {year} {2009})}\BibitemShut {NoStop}%
\bibitem [{\citenamefont {Das}\ \emph {et~al.}(2010)\citenamefont {Das},
  \citenamefont {Spaldin}, \citenamefont {Waghmare},\ and\ \citenamefont
  {Saha-Dasgupta}}]{PhysRevB.81.235112}%
  \BibitemOpen
  \bibfield  {author} {\bibinfo {author} {\bibfnamefont {H.}~\bibnamefont
  {Das}}, \bibinfo {author} {\bibfnamefont {N.~A.}\ \bibnamefont {Spaldin}},
  \bibinfo {author} {\bibfnamefont {U.~V.}\ \bibnamefont {Waghmare}},\ and\
  \bibinfo {author} {\bibfnamefont {T.}~\bibnamefont {Saha-Dasgupta}},\
  }\bibfield  {title} {\bibinfo {title} {Chemical control of polar behavior in
  bicomponent short-period superlattices},\ }\href
  {https://doi.org/10.1103/PhysRevB.81.235112} {\bibfield  {journal} {\bibinfo
  {journal} {Phys. Rev. B}\ }\textbf {\bibinfo {volume} {81}},\ \bibinfo
  {pages} {235112} (\bibinfo {year} {2010})}\BibitemShut {NoStop}%
\bibitem [{\citenamefont {Garc\'{\i}a-Fern\'andez}\ \emph
  {et~al.}(2013)\citenamefont {Garc\'{\i}a-Fern\'andez}, \citenamefont
  {Aguado-Puente},\ and\ \citenamefont {Junquera}}]{PhysRevB.87.085305}%
  \BibitemOpen
  \bibfield  {author} {\bibinfo {author} {\bibfnamefont {P.}~\bibnamefont
  {Garc\'{\i}a-Fern\'andez}}, \bibinfo {author} {\bibfnamefont
  {P.}~\bibnamefont {Aguado-Puente}},\ and\ \bibinfo {author} {\bibfnamefont
  {J.}~\bibnamefont {Junquera}},\ }\bibfield  {title} {\bibinfo {title}
  {Lattice screening of the polar catastrophe and hidden in-plane polarization
  in KNbO$_3$/BaTiO$_3$ interfaces},\ }\href
  {https://doi.org/10.1103/PhysRevB.87.085305} {\bibfield  {journal} {\bibinfo
  {journal} {Phys. Rev. B}\ }\textbf {\bibinfo {volume} {87}},\ \bibinfo
  {pages} {085305} (\bibinfo {year} {2013})}\BibitemShut {NoStop}%
\bibitem [{\citenamefont {Das}\ \emph {et~al.}(2011)\citenamefont {Das},
  \citenamefont {Waghmare},\ and\ \citenamefont
  {Saha-Dasgupta}}]{JApplPhys.109.066107}%
  \BibitemOpen
  \bibfield  {author} {\bibinfo {author} {\bibfnamefont {H.}~\bibnamefont
  {Das}}, \bibinfo {author} {\bibfnamefont {U.~V.}\ \bibnamefont {Waghmare}},\
  and\ \bibinfo {author} {\bibfnamefont {T.}~\bibnamefont {Saha-Dasgupta}},\
  }\bibfield  {title} {\bibinfo {title} {Piezoelectrics by design: A route
  through short-period perovskite superlattices},\ }\href
  {https://doi.org/10.1063/1.3561843} {\bibfield  {journal} {\bibinfo
  {journal} {J. Appl. Phys.}\ }\textbf {\bibinfo {volume} {109}},\ \bibinfo
  {pages} {066107} (\bibinfo {year} {2011})}\BibitemShut {NoStop}%
\bibitem [{\citenamefont {Zhu}(2018)}]{ChinPhysB.27.027701}%
  \BibitemOpen
  \bibfield  {author} {\bibinfo {author} {\bibfnamefont {Z.}~\bibnamefont
  {Zhu}},\ }\bibfield  {title} {\bibinfo {title} {First-principles study of
  polarization and piezoelectricity behavior in tetragonal PbTiO$_3$-based
  superlattices},\ }\href {https://doi.org/10.1088/1674-1056/27/2/027701}
  {\bibfield  {journal} {\bibinfo  {journal} {Chin. Phys. B}\ }\textbf
  {\bibinfo {volume} {27}},\ \bibinfo {pages} {027701} (\bibinfo {year}
  {2018})}\BibitemShut {NoStop}%
\bibitem [{\citenamefont {Park}\ and\ \citenamefont
  {Shrout}(1997)}]{JApplPhys.82.1804}%
  \BibitemOpen
  \bibfield  {author} {\bibinfo {author} {\bibfnamefont {S.-E.}\ \bibnamefont
  {Park}}\ and\ \bibinfo {author} {\bibfnamefont {T.~R.}\ \bibnamefont
  {Shrout}},\ }\bibfield  {title} {\bibinfo {title} {Ultrahigh strain and
  piezoelectric behavior in relaxor based ferroelectric single crystals},\
  }\href {https://doi.org/10.1063/1.365983} {\bibfield  {journal} {\bibinfo
  {journal} {J. Appl. Phys.}\ }\textbf {\bibinfo {volume} {82}},\ \bibinfo
  {pages} {1804} (\bibinfo {year} {1997})}\BibitemShut {NoStop}%
\bibitem [{\citenamefont {Guo}\ \emph {et~al.}(2000)\citenamefont {Guo},
  \citenamefont {Cross}, \citenamefont {Park}, \citenamefont {Noheda},
  \citenamefont {Cox},\ and\ \citenamefont {Shirane}}]{PhysRevLett.84.5423}%
  \BibitemOpen
  \bibfield  {author} {\bibinfo {author} {\bibfnamefont {R.}~\bibnamefont
  {Guo}}, \bibinfo {author} {\bibfnamefont {L.~E.}\ \bibnamefont {Cross}},
  \bibinfo {author} {\bibfnamefont {S.-E.}\ \bibnamefont {Park}}, \bibinfo
  {author} {\bibfnamefont {B.}~\bibnamefont {Noheda}}, \bibinfo {author}
  {\bibfnamefont {D.~E.}\ \bibnamefont {Cox}},\ and\ \bibinfo {author}
  {\bibfnamefont {G.}~\bibnamefont {Shirane}},\ }\bibfield  {title} {\bibinfo
  {title} {Origin of the high piezoelectric response in
  PbZr$_{1-x}$Ti$_x$O$_3$},\ }\href
  {https://doi.org/10.1103/PhysRevLett.84.5423} {\bibfield  {journal} {\bibinfo
   {journal} {Phys. Rev. Lett.}\ }\textbf {\bibinfo {volume} {84}},\ \bibinfo
  {pages} {5423} (\bibinfo {year} {2000})}\BibitemShut {NoStop}%
\bibitem [{\citenamefont {Rappe}\ \emph {et~al.}(1990)\citenamefont {Rappe},
  \citenamefont {Rabe}, \citenamefont {Kaxiras},\ and\ \citenamefont
  {Joannopoulos}}]{PhysRevB.41.1227}%
  \BibitemOpen
  \bibfield  {author} {\bibinfo {author} {\bibfnamefont {A.~M.}\ \bibnamefont
  {Rappe}}, \bibinfo {author} {\bibfnamefont {K.~M.}\ \bibnamefont {Rabe}},
  \bibinfo {author} {\bibfnamefont {E.}~\bibnamefont {Kaxiras}},\ and\ \bibinfo
  {author} {\bibfnamefont {J.~D.}\ \bibnamefont {Joannopoulos}},\ }\bibfield
  {title} {\bibinfo {title} {Optimized pseudopotentials},\ }\href
  {https://doi.org/10.1103/PhysRevB.41.1227} {\bibfield  {journal} {\bibinfo
  {journal} {Phys. Rev. B}\ }\textbf {\bibinfo {volume} {41}},\ \bibinfo
  {pages} {1227} (\bibinfo {year} {1990})}\BibitemShut {NoStop}%
\bibitem [{\citenamefont
  {Lebedev}(2009{\natexlab{b}})}]{PhysSolidState.51.362}%
  \BibitemOpen
  \bibfield  {author} {\bibinfo {author} {\bibfnamefont {A.~I.}\ \bibnamefont
  {Lebedev}},\ }\bibfield  {title} {\bibinfo {title} {Ab initio calculations of
  phonon spectra in \emph{A}TiO$_3$ perovskite crystals (\emph{A} = Ca, Sr, Ba, Ra, Cd, Zn,
  Mg, Ge, Sn, Pb)},\ }\href {https://doi.org/10.1134/S1063783409020279}
  {\bibfield  {journal} {\bibinfo  {journal} {Phys. Solid State}\ }\textbf
  {\bibinfo {volume} {51}},\ \bibinfo {pages} {362} (\bibinfo {year}
  {2009}{\natexlab{b}})}\BibitemShut {NoStop}%
\bibitem [{\citenamefont {Lebedev}(2016)}]{PhysSolidState.58.300}%
  \BibitemOpen
  \bibfield  {author} {\bibinfo {author} {\bibfnamefont {A.~I.}\ \bibnamefont
  {Lebedev}},\ }\bibfield  {title} {\bibinfo {title} {Phase transitions and
  metastable states in stressed SrTiO$_3$ films},\ }\href
  {https://doi.org/10.1134/S1063783416020190} {\bibfield  {journal} {\bibinfo
  {journal} {Phys. Solid State}\ }\textbf {\bibinfo {volume} {58}},\ \bibinfo
  {pages} {300} (\bibinfo {year} {2016})}\BibitemShut {NoStop}%
\bibitem [{\citenamefont {Yu}\ and\ \citenamefont
  {Krakauer}(1995)}]{PhysRevLett.74.4067}%
  \BibitemOpen
  \bibfield  {author} {\bibinfo {author} {\bibfnamefont {R.}~\bibnamefont
  {Yu}}\ and\ \bibinfo {author} {\bibfnamefont {H.}~\bibnamefont {Krakauer}},\
  }\bibfield  {title} {\bibinfo {title} {First-principles determination of
  chain-structure instability in KNbO$_3$},\ }\href
  {https://doi.org/10.1103/PhysRevLett.74.4067} {\bibfield  {journal} {\bibinfo
   {journal} {Phys. Rev. Lett.}\ }\textbf {\bibinfo {volume} {74}},\ \bibinfo
  {pages} {4067} (\bibinfo {year} {1995})}\BibitemShut {NoStop}%
\bibitem [{Note1()}]{Note1}%
  \BibitemOpen
  \bibinfo {note} {The numbers of irreducible representations used in this work
  follow their classification adopted at Bilbao Crystallographic Server~\cite
  {Bilbao}.}\BibitemShut {Stop}%
\bibitem [{\citenamefont {Pertsev}\ \emph {et~al.}(1998)\citenamefont
  {Pertsev}, \citenamefont {Zembilgotov},\ and\ \citenamefont
  {Tagantsev}}]{PhysRevLett.80.1988}%
  \BibitemOpen
  \bibfield  {author} {\bibinfo {author} {\bibfnamefont {N.~A.}\ \bibnamefont
  {Pertsev}}, \bibinfo {author} {\bibfnamefont {A.~G.}\ \bibnamefont
  {Zembilgotov}},\ and\ \bibinfo {author} {\bibfnamefont {A.~K.}\ \bibnamefont
  {Tagantsev}},\ }\bibfield  {title} {\bibinfo {title} {Effect of mechanical
  boundary conditions on phase diagrams of epitaxial ferroelectric thin
  films},\ }\href {https://doi.org/10.1103/PhysRevLett.80.1988} {\bibfield
  {journal} {\bibinfo  {journal} {Phys. Rev. Lett.}\ }\textbf {\bibinfo
  {volume} {80}},\ \bibinfo {pages} {1988} (\bibinfo {year}
  {1998})}\BibitemShut {NoStop}%
\bibitem [{\citenamefont {Di\'eguez}\ \emph {et~al.}(2004)\citenamefont
  {Di\'eguez}, \citenamefont {Tinte}, \citenamefont {Antons}, \citenamefont
  {Bungaro}, \citenamefont {Neaton}, \citenamefont {Rabe},\ and\ \citenamefont
  {Vanderbilt}}]{PhysRevB.69.212101}%
  \BibitemOpen
  \bibfield  {author} {\bibinfo {author} {\bibfnamefont {O.}~\bibnamefont
  {Di\'eguez}}, \bibinfo {author} {\bibfnamefont {S.}~\bibnamefont {Tinte}},
  \bibinfo {author} {\bibfnamefont {A.}~\bibnamefont {Antons}}, \bibinfo
  {author} {\bibfnamefont {C.}~\bibnamefont {Bungaro}}, \bibinfo {author}
  {\bibfnamefont {J.~B.}\ \bibnamefont {Neaton}}, \bibinfo {author}
  {\bibfnamefont {K.~M.}\ \bibnamefont {Rabe}},\ and\ \bibinfo {author}
  {\bibfnamefont {D.}~\bibnamefont {Vanderbilt}},\ }\bibfield  {title}
  {\bibinfo {title} {Ab initio study of the phase diagram of epitaxial
  BaTiO$_3$},\ }\href {https://doi.org/10.1103/PhysRevB.69.212101} {\bibfield
  {journal} {\bibinfo  {journal} {Phys. Rev. B}\ }\textbf {\bibinfo {volume}
  {69}},\ \bibinfo {pages} {212101} (\bibinfo {year} {2004})}\BibitemShut
  {NoStop}%
\bibitem [{Note2()}]{Note2}%
  \BibitemOpen
  \bibinfo {note} {The periodicity of a superlattice assumes that the electric
  field strengths $E_1$ and $E_2$ in its layers satisfy the condition $E_1x_1 +
  E_2x_2 = 0$, where $x_1$ and $x_2$ are the thicknesses of individual layers.
  This condition combined with the equation $(P_1 + \epsilon _1 E_1) - (P_2 +
  \epsilon _2 E_2) = \Delta P$, in which $P_1$, $P_2$, $\epsilon _1$, and
  $\epsilon _2$ are the spontaneous polarizations and dielectric constants in
  two layers, gives a solution to this problem in the linear
  approximation.}\BibitemShut {Stop}%
\bibitem [{\citenamefont {Zhu}\ \emph {et~al.}(2016)\citenamefont {Zhu},
  \citenamefont {Wang},\ and\ \citenamefont {Fu}}]{ChinPhysLett.33.026302}%
  \BibitemOpen
  \bibfield  {author} {\bibinfo {author} {\bibfnamefont {Z.-Y.}\ \bibnamefont
  {Zhu}}, \bibinfo {author} {\bibfnamefont {S.-Q.}\ \bibnamefont {Wang}},\ and\
  \bibinfo {author} {\bibfnamefont {Y.-M.}\ \bibnamefont {Fu}},\ }\bibfield
  {title} {\bibinfo {title} {First-principles study of properties of strained
  PbTiO$_3$/KTaO$_3$ superlattice},\ }\href
  {https://doi.org/10.1088/0256-307X/33/2/026302} {\bibfield  {journal}
  {\bibinfo  {journal} {Chin. Phys. Lett.}\ }\textbf {\bibinfo {volume} {33}},\
  \bibinfo {pages} {026302} (\bibinfo {year} {2016})}\BibitemShut {NoStop}%
\bibitem [{Bil()}]{Bilbao}%
  \BibitemOpen
  \href {http://www.cryst.ehu.es/} {\bibinfo {title} {Bilbao crystallographic
  server}},\ \bibinfo {note} {\url{http://www.cryst.ehu.es/}}\BibitemShut
  {NoStop}%
\end{thebibliography}

%
% ****** End of file apstemplate.tex ******
%apsrev4-2.bst 2019-01-14 (MD) hand-edited version of apsrev4-1.bst
%Control: key (0)
%Control: author (8) initials jnrlst
%Control: editor formatted (1) identically to author
%Control: production of article title (0) allowed
%Control: page (0) single
%Control: year (1) truncated
%Control: production of eprint (0) enabled
\providecommand{\BIBYu}{Yu}

\end{document}